\documentclass[apjl]{emulateapj}
\def\C10{C10b}
\def\be{\begin{equation}} 
\def\ee{\end{equation}}

\def\gsim{\lower.5ex\hbox{\gtsima}} 
\def\lsim{\lower.5ex\hbox{\ltsima}} \def\gtsima{$\; \buildrel > \over 
\sim \;$} \def\ltsima{$\; \buildrel < \over \sim \;$} \def\prosima{$\; 
\buildrel \propto \over \sim \;$} \def\gsim{\lower.5ex\hbox{\gtsima}} 
\def\lsim{\lower.5ex\hbox{\ltsima}} 
\def\simgt{\lower.5ex\hbox{\gtsima}} 
\def\simlt{\lower.5ex\hbox{\ltsima}} 
\def\simpr{\lower.5ex\hbox{\prosima}}   
  
 \def\gtsima{$\; \buildrel > \over \sim \;$} 
\def\ltsima{$\; \buildrel < \over \sim \;$} 
\def\gsim{\lower.5ex\hbox{\gtsima}} 
\def\lsim{\lower.5ex\hbox{\ltsima}} 
\def\simgt{\lower.5ex\hbox{\gtsima}} 
\def\simlt{\lower.5ex\hbox{\ltsima}} 
\def\simpr{\lower.5ex\hbox{\prosima}}

\def\E3{{\cal E}_{\rm g}^{III}}

\def\avchi{$\langle \chi_{HI} \rangle$}
\def\hchi{$ \chi_{HI} $}
\shorttitle{First observational support for overlapping reionized bubbles generated by a galaxy overdensity}
\shortauthors{M. Castellano et al.}

\begin{document}

\title{First observational support for overlapping reionized bubbles generated by a galaxy overdensity}
\author{M. Castellano\altaffilmark{1}, P. Dayal\altaffilmark{2,3}, L. Pentericci\altaffilmark{1}, A. Fontana\altaffilmark{1}, A. Hutter\altaffilmark{4}, G. Brammer\altaffilmark{5}, E. Merlin\altaffilmark{1}, A. Grazian\altaffilmark{1}, S. Pilo\altaffilmark{1}, R. Amorin\altaffilmark{1}, S. Cristiani\altaffilmark{6,7}, M. Dickinson\altaffilmark{8}, A. Ferrara\altaffilmark{9}, S. Gallerani\altaffilmark{9}, E. Giallongo\altaffilmark{1}, M. Giavalisco\altaffilmark{10},  L. Guaita\altaffilmark{1}, A. Koekemoer\altaffilmark{5}, R. Maiolino\altaffilmark{11,12}, D. Paris\altaffilmark{1}, P. Santini\altaffilmark{1}, L. Vallini\altaffilmark{9,13}, E. Vanzella\altaffilmark{14}, J. Wagg\altaffilmark{15}}

\altaffiltext{1}{INAF - Osservatorio Astronomico di Roma, Via Frascati 33, I - 00040 Monte Porzio Catone (RM), Italy}
\altaffiltext{2}{Institute for Computational Cosmology, Department of Physics, University of Durham, South Road, Durham DH1 3LE, UK}
\altaffiltext{3}{Kapteyn Astronomical Institute, University of Groningen, P.O. Box 800, 9700, AV Groningen, The Netherlands}
\altaffiltext{4}{Swinburne University of Technology, Hawthorn, VIC 3122, Australia }
\altaffiltext{5}{Space Telescope Science Institute, 3700 San Martin Drive, Baltimore, MD 21218, USA}
\altaffiltext{6}{INAF - Osservatorio Astronomico di Trieste, Via G. B. Tiepolo 11, I-34143 Trieste, Italy}
\altaffiltext{7}{INFN - National Institute for Nuclear Physics, via Valerio 2, I-34127,  Trieste, Italy}
\altaffiltext{8}{National Optical Astronomy Observatories, Tucson, AZ 85719, USA}
\altaffiltext{9}{Scuola Normale Superiore, Piazza dei Cavalieri 7, I-56126 Pisa, Italy}
\altaffiltext{10}{Astronomy Department, University of Massachusetts, Amherst, MA 01003, USA}
\altaffiltext{11}{Cavendish Laboratory, University of Cambridge, 19 J. J. Thomson Ave, Cambridge CB3 0HE, UK }
\altaffiltext{12}{Kavli Institute for Cosmology, University of Cambridge, Madingley Road, Cambridge CB3 0HA, UK}
\altaffiltext{13}{Dipartimento di Fisica e Astronomia, Universit\`{a} di Bologna, viale Berti Pichat 6/2, I-40127 Bologna, Italy}
\altaffiltext{14}{INAF - Osservatorio Astronomico di Bologna, Via Ranzani 1, I - 40127, Bologna, Italy}
\altaffiltext{15}{Square Kilometre Array Organization, Jodrell Bank Observatory, Lower Withington, Macclesfield, Cheshire SK11 9DL, UK}
\email{marco.castellano\char64oa-roma.inaf.it}


\begin{abstract}
We present the analysis of deep HST multi-band imaging of the BDF field specifically designed to identify faint companions around two of the few Ly$\alpha$ emitting galaxies spectroscopically confirmed at z$\sim$7 \citep{Vanzella2011}. Although separated by only 4.4 proper Mpc these galaxies cannot generate HII regions large enough to explain visibility of their Ly$\alpha$ line, thus requiring a population of fainter ionizing sources in their vicinity. We use deep HST and VLT-Hawk-I data to select z$\sim$7 Lyman break galaxies around the emitters. We select 6 new robust z$\sim$7 LBGs at Y$\sim$26.5-27.5 whose average spectral energy distribution is consistent with the objects being at the redshift of the close-by Ly$\alpha$ emitters. The resulting number density of z$\sim$7 LBGs in the BDF field is a factor$\sim$3-4 higher than expected in random pointings of the same size. We compare these findings with cosmological hydrodynamic plus radiative transfer simulations of a universe with a half neutral IGM: we find that indeed Ly$\alpha$ emitter pairs are only found in completely ionized regions characterized by significant LBG overdensities. Our findings match the theoretical prediction that the first ionization fronts are generated within significant galaxy overdensities and support a scenario where faint, ``normal'' star-forming galaxies are responsible for reionization.
\end{abstract}



\section{Introduction}
In recent years, spectroscopic follow-up campaigns of z$\sim$7 Lyman break galaxies (LBG) have enabled a direct investigation of the timeline of the reionization process by studying the redshift evolution of the Ly$\alpha$ fraction in LBGs \citep{Stark2010}, which is expected to fall-off when the IGM becomes significantly neutral and Ly$\alpha$ emission is attenuated \citep{Dijkstra2015}.  A substantial decrease of the Ly$\alpha$ fraction between z$\sim$6 and z$\sim$7 has been measured \citep{Fontana2010,Pentericci2011,Schenker2012,Caruana2012,Ono2012}, with latest data favoring a scenario with a change of the neutral hydrogen fraction of $\Delta \chi_{HI}\sim$0.5 in a redshift interval $\Delta$z=1, and a patchy reionization process \citep{Treu2012,Pentericci2014}.
\citet{Bouwens2015b} have shown that, under plausible assumptions on the properties of star-forming galaxies, the evolution of their UV luminosity density can explain the reionization timeline estimated from spectroscopic data and other probes. However, significant uncertainties remain: a direct evidence of the connection between galaxies and reionization as well as an explanation of the patchiness found from spectroscopy are still missing. Among the 8 independent lines of sight analysed by \citet{Pentericci2014}, the Bremer Deep Field (BDF) stands out as a peculiar area in the z$\sim$7 Universe. The BDF is the only field where two close-by bright Ly$\alpha$-emitting galaxies, BDF-3299 (z=7.109) and BDF-521 (z=7.008) have been found \citep{Vanzella2011}. These two L$\sim$L$^*$ objects, originally selected as z-dropout candidates \citep[][C10b hereafter]{Castellano2010b}, show Ly$\alpha$ equivalent width $>$ 50\AA~and are separated by a distance of only 4.4 proper Mpc (pMpc). The detection of bright Ly$\alpha$ emission from them can be explained by these sources being embedded in an HII region that allows Ly$\alpha$ photons to redshift far away from the line center before they reach the almost neutral IGM. \citet{Vanzella2011} has compared the size of the HII region these galaxies can build \citep{Loeb2005} to the minimum ionized radius allowing their Ly$\alpha$ photons to escape \citep{Wyithe2005}, finding that they both cannot generate HII regions large enough to explain the visibility of their lines even if a maximum Lyman continuum escape fraction f$_{esc}$=1 is assumed. Instead, the visibility of their Ly$\alpha$ can be explained by the presence of additional ionizing sources in their vicinity \citep{Dayal2009}. Unfortunately, the data available at that time did not allow to select galaxies fainter than the two emitters to constrain this hypothesis. In this paper we present the analysis of deep HST observations of the BDF region (cycle 22 program 13688, P.I. M. Castellano) specifically designed to detect possible fainter companions of BDF-3299 and BDF-521. 
Throughout the paper, magnitudes are in the AB system, and we adopt the $\Lambda$-CDM concordance model ($H_0=70km/s/Mpc$, $\Omega_M=0.3$, and $\Omega_{\Lambda}=0.7$).

\section{Dataset and z$\sim$7 LBG selection}\label{dataset}

The HST observations of the BDF field consists of two pointings including BDF-3299 and BDF-521 and designed to cover the largest possible portion of the region between them. Each pointing has been observed with WFC3 in the Y105 filter (3 orbits) and with ACS filters I814 (3 orbits) and V606 (2 orbits). The HST images were processed as described in detail by \citet{Momcheva2015}, taking into account time-variable backgrounds and high-altitude atmospheric line emission \citep{Brammer2014}. The individual exposures in each band were registered and combined with \verb|DrizzlePac| \citep{Gonzaga2012} into mosaics with 0.06 arcsec pixel scale. Coverage redward of the Y105 band is needed to separate LBGs from low-redshift red interlopers. To this aim we exploit the J and K band HAWK-I data (C10b) to build a J+K mosaic as weighted average of the two datasets after having matched the Ks band PSF to the J band one. We include in the analysis our reduction of the J125 and H160 observations of the BDF-521 field acquired under program 12487 \citep[PI X. Fan,][]{Cai2015} that partially overlap with our Y105 data. We detect sources in the Y105 band with \verb|SExtractor| \citep{Bertin1996} and use the relevant \verb|FLUX_AUTO| value as total flux in this band. We PSF-match the other HST bands to the Y105 one through appropriate convolution kernels and then use \verb|SExtractor| in dual-mode to obtain total magnitudes in each band by scaling total Y105 flux on the basis of the relevant isophotal colour. To measure photometry from the J+K mosaic whose resolution (0.55 arcsec PSF FWHM) is signficantly lower than the Y105 one (0.19 arcsec) we perform double-pass \verb|T-PHOT| runs using source cutouts from the Y105 image as reference high-resolution templates \citep{Merlin2015}.

The resulting magnitude limits in 2$\times$FWHM apertures are: V606=30.6, I814=30.4, (S/N=1), Y105=28.0 (S/N=10), (J+K)=27.6 (S/N=1).
\begin{figure}[!ht]
   \centering
   \resizebox{\hsize}{!}{\includegraphics[]{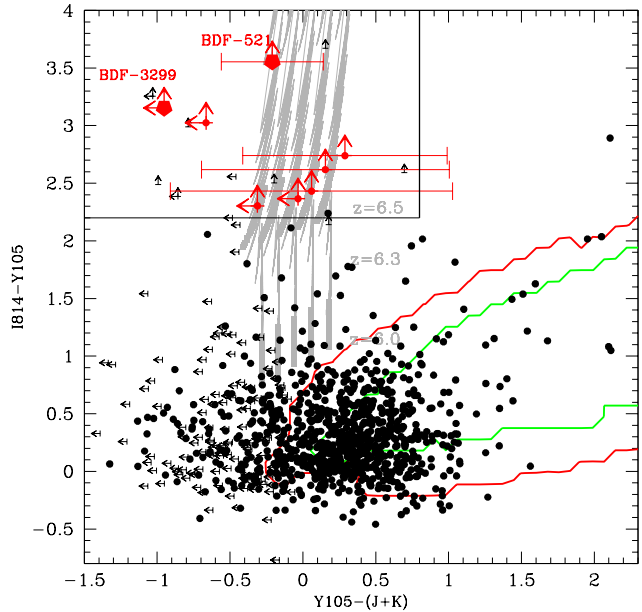}}
   \caption{I814-Y105 vs. Y105-(J+K) diagram of the objects in the HST-BDF catalogues. The z$\sim$7 LBG selection region is enclosed by the box: the selected candidates are marked in red, with BDF-521 and BDF-3299 indicated by filled pentagons. Grey lines show the position of star-forming BC03 models at z$>$6.0 (see labels) with 0.0$<$E(B-V)$<$0.2 (from left to right), age=100Myr, constant SFH, Z=0.2Z$_{\odot}$. Red and green contours show respectively the position of passive and dusty BC03 templates at z$<$4.}\label{fig_seldiag}%
\end{figure}

We use templates from the \citet{Bruzual2003} (BC03 hereafter) library including both high- and low-z galaxies to define the following LBG selection window (Fig.~\ref{fig_seldiag}): 
\begin{eqnarray*}
&&(S/N(I_{814})<1) \wedge (I_{814}-Y_{105}>2.2)\\
&&Y_{105}-(J+K)<0.8\\
&&(S/N(Y_{105})>10) \wedge (S/N(V_{606}))<1,
\end{eqnarray*} 
where the signal-to-noise ratio in the Y105 band is associated to the total flux and relevant uncertainty in this band, and S/N in the ACS bands are measured in 2$\times$FWHM apertures. The strict non-detection in the I814 band, plus the requirement on I814-Y105$>$2.2 (that is met at z$>6.5$) ensures that we are selecting objects in the standard z-dropout redshift range.  We also apply additional criteria to include only objects with robust photometry. These have been defined with the procedure outlined in \citet{Castellano2010} based on the analysis of a `negative' detection image, ensuring that no spurious detections are expected for S/N(Y105)$>$10 samples. To avoid noisy areas we retrict the analysis to regions where the r.m.s of the images is less than $\sim$ 1.5 times the lowest value. The resulting total area used for the analysis is 3.94 and 3.82 sq. arcmins in the BDF-521 and BDF-3299 pointings respectively. We also require an optimal \verb|SExtractor| extraction flag (\verb|FLAG|=0), and an isophotal area \verb|ISOAREA|$>$18 pixels (equal to 2$\times$ the area of one FWHM of the Y105 PSF) to avoid detections possibly due to residual cosmic-rays or hot pixels.

Our selection criteria yield 4 LBG candidates in each BDF pointing (Fig.~\ref{fig_seldiag}). Reassuringly both BDF-521 and BDF-3299 are included in our selected sample together with 6 additional LBGs at Y105$\sim$26.5-27.5. Among the six newly discovered LBGs, three sources show  a marginal detection in the (J+K) mosaic at S/N$\sim$1-1.5. Two of the sources in the BDF-521 pointing (one of them undetected in J+K) are also included in the area of the WFC3 observations from \citet{Cai2015} and are detected at S/N$\sim$2-8 in both the J125 and H160 bands with colours expected for z$\sim$7 LBGs.  All our BDF candidates are extended with half-light radius in the range Rh$\sim$0.09-0.17 arcsec, consistent with expectations for z$\sim$7 galaxies in the same magnitude range \citep{Grazian2012}. In fact, they are classified with \verb|CLASS_STAR|$<$0.9 such that we can exclude that cool stars and transient objects contaminate our sample, since this kind of contaminants are expected to have a higher stellarity index at S/N$>$10 \citep{Bouwens2015}. We compute the photometric redshift of the sources by fitting their photometry with our $\chi^2$ minimization code \citep{Fontana2000} using a library of BC03 templates at z=0-8 \citep[including line emission as in][]{Schaerer2009,Castellano2014}. We find that all sources have best-fit solutions at high-z (z$\sim$6.8-7.4) and consistent, within the  1$\sigma$ uncertainty,  with the spectroscopic redshift of the two emitters.

\subsection{Test of the LBG selection criteria on the HUDF}\label{HUDF}
As a test of our selection criteria we apply them to a real case by degrading HUDF V606, I814, Y105, J125 and Ks images \citep{Koekemoer2013,Fontana2014} to the depth and resolution of the BDF dataset. This is particularly interesting to test the efficiency of the J+K data in separating red low-z interlopers (expected at S/N$\gtrsim$3) from LBGs (close to the detection limit at S/N=1). In practice, we add Gaussian noise to the V606, I814, Y105 and Ks images to match the depth of the BDF observations, and both smooth and decrease depth of the J125 image to match the J-HAWKI mosaic of the BDF. We build a (J+K) mosaic and extract catalogue and photometry in the same way as in the BDF case. Our selection criteria yield only one candidate: the LBG G2\_1408 from \citet{Castellano2010}, well known in the literature being selected by all  analysis of the HUDF field \citep[see][]{Vanzella2014}. The other z$\sim$7 LBGs in the HUDF from \citet{Bouwens2015} have Y105$\gtrsim$27.7 \citep[photometry from][]{Guo2013} and are thus not expected as S/N$>$10 sources in our data.  Most importantly, we find that no z$<$6.5 objects are scattered into our LBG selection window. In addition, no spurious detections are found, consistently with our finding from the ``negative image'' test.

\subsection{Stacking of the newly identified LBGs}

\begin{figure}[!ht]
   \centering
   \resizebox{\hsize}{!}{\includegraphics[width=7.5cm]{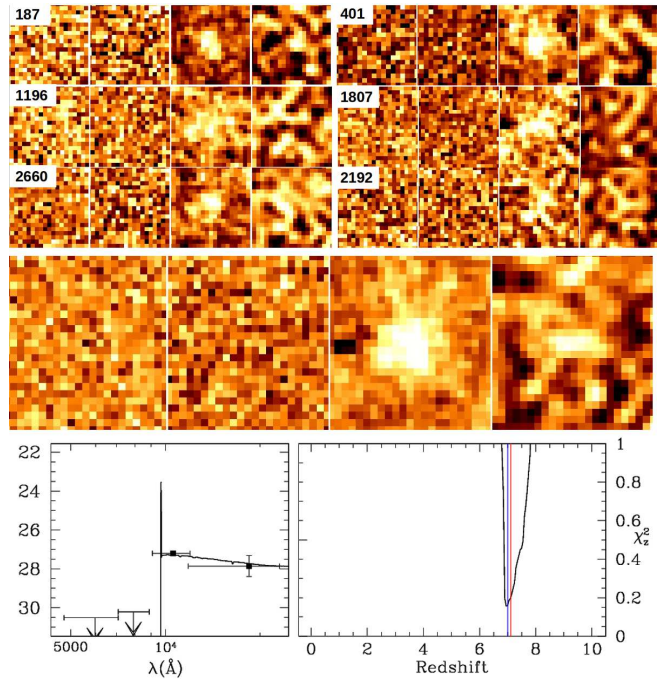}}
   %\resizebox{\hsize}{!}{\includegraphics[width=7.5cm]{fig2.eps}}
  \caption{\textbf{Top}: thumbnails (1 arcsec side) of the newly detected LBGs in the BDF-521 (left column) and BDF-3299 (right column) HST pointings in, from left to right, V606, I814, Y105 and (J+K) bands. \textbf{Middle}: stacked thumbnails. \textbf{Bottom}: best-fit spectral energy distribution of the stacked object (left panel) and $\chi^2$ as a function of redshift (right panel, red and blue line mark the spectroscopic redshifts of BDF-3299 and BDF-521 respectively).}\label{fig_stack}%
\end{figure}

To further verify that the 6 newly discovered LBGs are genuine z$\sim$7 sources we build stacked (weighted average) images in the different bands enabling a tighter constraint on the Lyman break. We extract photometry from the stacked images with \verb|SExtractor| and \verb|T-PHOT|. We find non-detections in both the V606 and I814, corresponding to a limiting total magnitude $>$30.2 and to a color I814-Y105$>$3, and a S/N$\sim$2 detection in the J+K one. The stacked thumbnails and the resulting SED are shown in Fig.~\ref{fig_stack}. We find a best-fit photometric redshift z=6.95, with P($\chi^2$)$>$32\% solutions constrained to the range z=6.8-7.8, as expected from the sharp I814-Y105 drop. The spectroscopic redshifts of BDF-521 and BDF-3299 are close to the best-fit solution. 
\begin{figure}[ht]
   \centering
   \resizebox{\hsize}{!}{\includegraphics[]{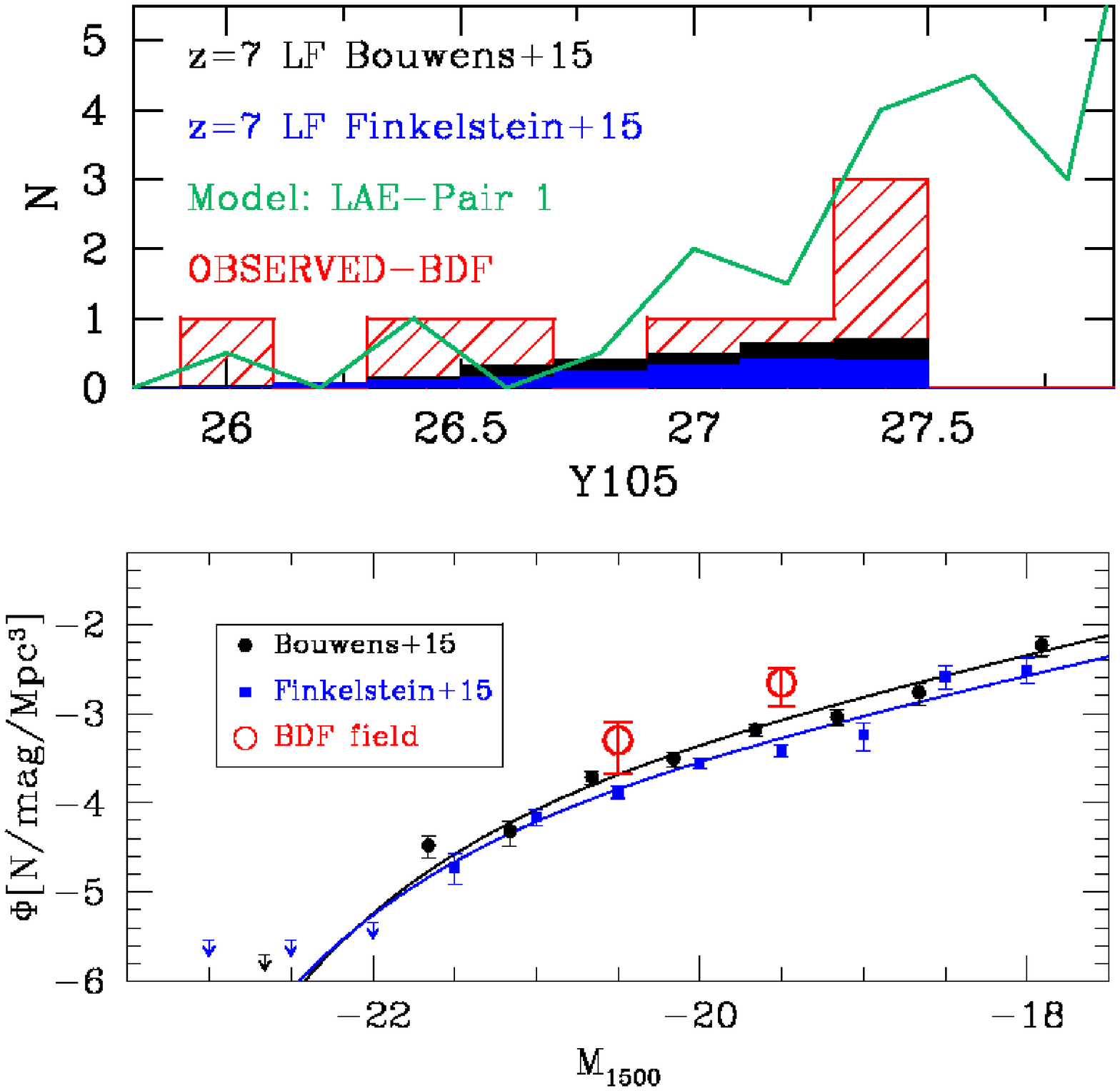}}
   \caption{\textbf{Top}: Observed number counts of z$\sim$7 LBGs in the BDF field (red histogram) compared to the number counts expected from z$\sim$7 LFs, and to number counts in the region of ``Pair 1'' in our cosmological simulations (green line).
\textbf{Bottom}: Stepwise z$\sim$7 UV LF in the BDF field (red open circles) and the average one from wide surveys \citep[][black circles and blue squares respectively]{Bouwens2015,Finkelstein2015}. The best-fit z$\sim$7 LFs from \citet{Bouwens2015} and \citet{Finkelstein2015} are shown as black and blue lines respectively.}\label{fig_counts}%
\end{figure}
\section{An overdensity of LBGs in the BDF}\label{discussion}
We use extensive simulations for a detailed assessment of the LBG number density observed in the BDF field, and in particular to compare it to expectations from our current knowledge of the z$\sim$7 UV LF.

\subsection{Expected number counts}

We first generate a library of BC03 models at 6.0$<$z$<$8, with constant SFH, age from 10Myr to the age of the universe at the relevant redshift, 0.0$<$E(B-V)$<$0.2 \citep[extinction law from][]{Calzetti2000}, and metallicities Z=0.02,0.2,1.0 Z$_\odot$. We randomly extract from it a reference catalogue with 75000 sources whose fluxes are renormalized so to follow a constant distribution at -21.5$<$M$_{1500}<$-18.5. This  catalogue is used as input for simulations mimicking the survey properties. Observed magnitudes in the V606, I814 and Y105 bands are obtained by inserting mock galaxies in the real images and assembling the catalogue as in the real case (see, e.g., C10b). We assume galaxy shapes to be disk-like with Rh randomly drawn from the distribution by \citet{Grazian2012}. Imaging simulations are exceedingly time consuming in the case of \verb|T-PHOT| photometry. We thus include J+K ``observed'' magnitudes following the technique described in \citet{Castellano2012}, namely we perturb the template fluxes through Monte Carlo simulations designed to reproduce both the average and the scatter of the (total) S/N vs. magnitude relation in the observed J+K datasets. We then randomly extract from our reference library galaxy populations following the latest estimates of the z$\sim$7 UV LF from \citet{Bouwens2015} and \citet{Finkelstein2015}. We select objects in the same way as for the observed datasets and the resulting number counts are scaled to the observed area. Simulations and LBG selection are performed separately in the two pointings to take into account small differences in depth and coverage between them. We find that a total of $N_{exp}$=1.8-2.9 objects are expected in our HST survey of the BDF on the basis of the LFs from \citet{Finkelstein2015} and \citet{Bouwens2015} respectively. The probability of detecting eight sources in our field out of a Poisson distribution with $N_{exp}$=1.8-2.9, is only 0.045-0.68\%, while the estimated cosmic variance\footnote{computed with the Cosmic Variance Calculator \citep{Trenti2008}, \texttt{http://casa.colorado.edu/~trenti/CosmicVariance.html}} is $\sim$0.5$\cdot N_{exp}$.  \textit{This analysis confirms that the BDF field with its eight detected LBGs is a factor of $\sim$3-4 overdense at z$\sim$7 with respect to the average galaxy number density at these redshifts (Fig.~\ref{fig_counts})}. 
\subsection{Comoving number density}
We recast our analysis into a comparison between the observed z$\sim$7 LF in the BDF field and the average one from wide surveys. We perform a binned estimate of the comoving number density through a stepwise method (e.g. C10b). In practice, we assume the objects to be at z=7 to convert Y band magnitudes into M$_{1500}$ rest-frame, and measure the number density in two bins as N$_{obj}$/V$_{eff}$, where V$_{eff}$ is the effective volume probed by the survey at z=6.5-7.5 as estimated from the simulations. The results are shown in the bottom panel of Fig.~\ref{fig_counts} and confirm an excess of galaxies in the BDF field both at bright and faint magnitudes, although in the brightest bin (3 sources) the BDF number density remains consistent with the LF by \citet{Bouwens2015} within the Poissonian uncertainty. We underline that this is a conservative estimate since it has been obtained by considering the full effective volume probed by our survey. As an example, the real comoving number density in the BDF is $\sim$15 times higher than the average in case the six additional LBGs detected with HST data are physically associated with BDF-521 and BDF-3299 in a unique structure at z$\sim$7.1 with $\Delta z$=0.25.
\begin{figure}[ht]
   \centering
  \includegraphics[width=7.cm]{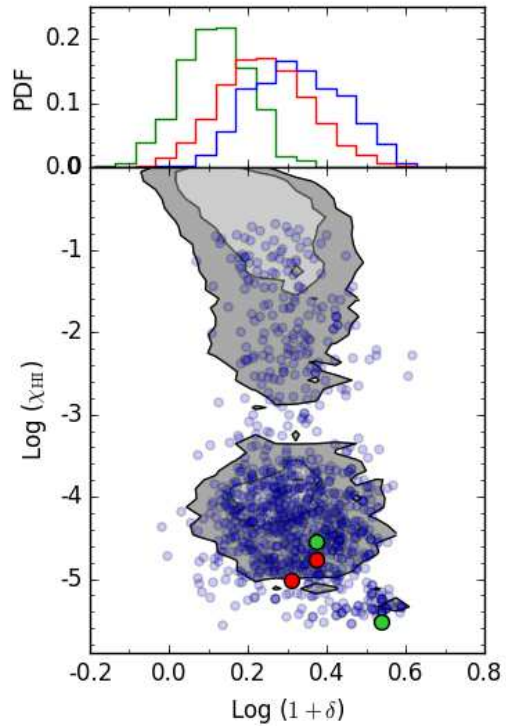}
   \caption{\textbf{Bottom panel}: hydrogen neutral fraction versus galaxy overdensity in our cosmological simulations. Grey contours show the region occupied by LBGs, blue circles mark the position of LAEs. The LAEs in ``clustered pairs'' are shown in green (Pair 1) and red (Pair 2). \textbf{Top panel}: density distribution of the LAEs (blue) and of the LBGs in regions with hydrogen neutral fraction above (red) and below (green) the average \hchi$=$0.5.}\label{fig_model}%
 \end{figure}
\section{Discussion}
The BDF field, showing both close-by emitters and a galaxy overdensity, displays all the properties of an early reionized region expected in theoretical models that postulate a dependence between galaxy density and the reionization timeline, with overdense regions being the first to become reionized ``bubbles'' \citep{McQuinn2007,Wyithe2007,Dayal2009,Iliev2014}. To test this scenario we should eventually prove that the BDF environment is not only overdense with respect to the average, but also to the galaxy number density around other Y105$\sim$26 galaxies at the same redshift, that show Ly$\alpha$ emission only in $\sim$15\% of the cases. However, the only other field in \citet{Pentericci2014} where imaging data of comparable depth are available is GOODS-South: interestingly, we verified that all the 3 galaxies with Y105$<$26.5 in this field lack both Ly$\alpha$ and any significant overdensity around them. Given the limitation of available data in constraining the proposed scenario, we carry out a more in-depth evaluation through our cosmological simulations \citep{hutter2014, hutter2015} that model both the overall reionization process and the UV and Ly$\alpha$ emission properties of z$\simeq 7$ galaxies. 
\begin{figure}[ht]
   \centering
  \includegraphics[width=9.0cm]{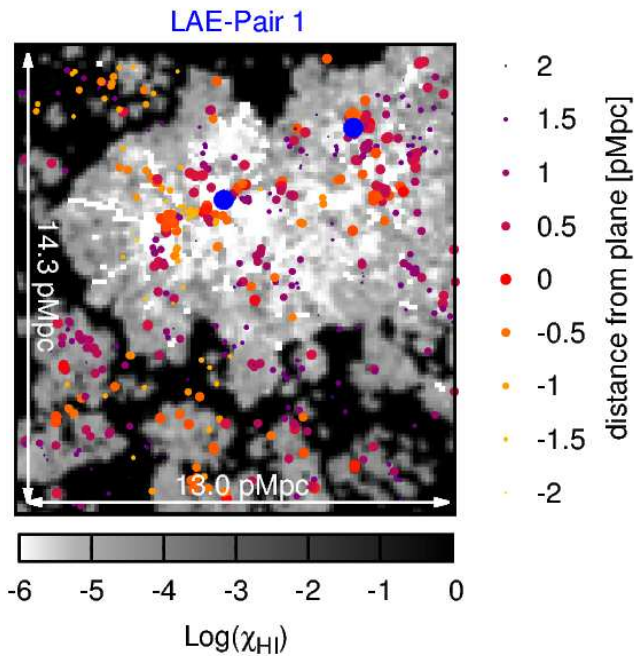}
   \caption{``Pair 1'' region in the model with main LAEs (blue circles) and their companion LBGs shown as circles with color and dimension indicating the distance from the displayed plane.}\label{fig_snapshot}%
\end{figure}
\subsection{Comparison with a SPH theoretical model}
The model couples cosmological SPH simulations run using \verb|GADGET-2| \citep{Springel2005} with a radiative transfer code \citep[pCRASH,][]{Partl2011} and a dust model. We explore a wide range of $f_{esc}$ ranging between 5-95\%. For each of these values, we couple the z$\sim$7 simulation snapshot with \verb|pCRASH|, starting from a fully neutral IGM and ending the runs once the IGM is fully ionized. We look for galaxies resembling BDF-521 and BDF-3299 in terms of Ly$\alpha$ and UV luminosity in a snapshot with average neutral faction \avchi$\simeq0.5$ consistent with current estimates at z$\sim$7 \citep{Bouwens2015,Mitra2015}. We take $f_{esc}=0.5$ as reference ionizing escape fraction. However, our analysis is unaffected by the $f_{esc}$  value adopted: as shown in \citet{hutter2014, hutter2015} the visibility of LAEs is governed by three degenerate quantities, $f_{esc}$, \avchi~and the dust absorption; at a given  \avchi, a lower (higher) $f_{esc}$ could be compensated by a lower (higher) dust absorption mostly leaving both the reionization topology and the fraction of LAEs unchanged. Our simulation contains 75 sub-volumes equal to the observed one: we find 7 (10) LAEs that match the Ly$\alpha$ and UV ranges of BDF-521 (BDF-3299), but only {\it two pairs} that are at comparable distance ($\sim 4$pMpc) - implying that the existence of such ``clustered pairs'' is rare, with a probability of only about 2.6\%. This value is roughly consistent with current findings, the BDF emitters being the only pair found among 68 z-dropout galaxies surveyed by \citet{Pentericci2014}. We compute galaxy density and average neutral fraction in regions equivalent to one HST pointing around the two pairs and compare them with values measured around isolated LAEs and normal LBGs in the model (Fig.~\ref{fig_model}). We find that both ``clustered pairs'' lie in \textit{highly ionized regions} and are characterized by a \textit{significant clustering} of LBGs in their surroundings. In general, there is an evident relation between neutral hydrogen fraction and galaxy density at \hchi$>$0.1. Model LAEs are found at \hchi$\lesssim$0.1, with clustered LAEs being embedded in overdense regions with a very low neutral HI fraction of log(\hchi)$\sim$-5. By inspecting the LBG populations surrounding the ``clustered pairs'' we find that the first pair shows LBG number counts very similar to the BDF observed ones (green line in Fig.~\ref{fig_counts}), while the second group show a LBG overdensity at Y$\gtrsim$28, fainter than the BDF limiting magnitude. A snapshot of Pair 1 from the model is shown in  Fig.~\ref{fig_snapshot}: the LAEs and companion LBGs lie in an ionized region with a radius of about $\simeq 5$pMpc, where \hchi~increases from 10$^{-6}$ to 10$^{-3}$ up to a sharp transition boundary with the mostly neutral IGM. This lends support to our suggestion that the BDF field hosts an early reionized region pointing to a connection between galaxy clustering and the reionization timeline.

\section{Summary and Conclusions}\label{summary}
The analysis of dedicated HST observations has shown that the BDF field, where we previously detected a unique pair of Ly-$\alpha$ emitting galaxies at z$\sim$7 \citep{Vanzella2011}, presents a number density of z$\sim$7 LBGs which is larger by a factor $\sim$3-4 than the average one.
A comparison between observations and cosmological simulations shows that the BDF likely hosts overlapping reionized regions with a very low neutral fraction (\hchi$<$10$^{-3}$) embedded in a half neutral IGM. Our findings match the expectation that overdense regions are the first to become reionized, and suggest that source clustering is a likely explanation for the inhomogeneity of reionization measured from spectroscopy \citep{Pentericci2014}. Finally, the consistency with model predictions on the relation between clustering and neutral hydrogen fraction, adds further evidence to a scenario where faint star-forming galaxies play a major role in reionization \citep{Bouwens2015b}. This picture clearly highlights the potentiality of going beyond the standard approach based on volume-averaged quantities, and investigate instead the properties of the ionizing and Ly$\alpha$-emitting sources as a function of different environments in order to constrain the unfolding of the reionization epoch.

\acknowledgments
Based on observations made with the NASA/ESA Hubble Space Telescope, obtained at the Space Telescope Science Institute, which is operated by the Association of Universities for Research in Astronomy, Inc., under NASA contract NAS 5-26555. These observations are associated with program \#13688. The research leading to these results has received funding from the European Union Seventh Framework Programme (FP7/2007-2013) under grant agreement n° 312725.


\begin{thebibliography}{41}
\expandafter\ifx\csname natexlab\endcsname\relax\def\natexlab#1{#1}\fi

\bibitem[{{Bertin} \& {Arnouts}(1996)}]{Bertin1996}
{Bertin}, E. \& {Arnouts}, S. 1996, \aaps, 117, 393

\bibitem[{{Bouwens} {et~al.}(2015{\natexlab{a}}){Bouwens}, {Illingworth},
  {Oesch}, {Caruana}, {Holwerda}, {Smit}, \& {Wilkins}}]{Bouwens2015b}
{Bouwens}, R.~J., {Illingworth}, G.~D., {Oesch}, P.~A., {et~al.}
  2015{\natexlab{a}}, \apj, 811, 140

\bibitem[{{Bouwens} {et~al.}(2015{\natexlab{b}}){Bouwens}, {Illingworth},
  {Oesch}, {Trenti}, {Labb{\'e}}, {Bradley}, {Carollo}, {van Dokkum},
  {Gonzalez}, {Holwerda}, {Franx}, {Spitler}, {Smit}, \& {Magee}}]{Bouwens2015}
{Bouwens}, R.~J., {Illingworth}, G.~D., {Oesch}, P.~A., {et~al.}
  2015{\natexlab{b}}, \apj, 803, 34

\bibitem[{{Brammer} {et~al.}(2014){Brammer}, {Pirzkal}, {McCullough}, \&
  {MacKenty}}]{Brammer2014}
{Brammer}, G., {Pirzkal}, N., {McCullough}, P., \& {MacKenty}, J. 2014,
  {Time-varying Excess Earth-glow Backgrounds in the WFC3/IR Channel}, Tech.
  rep.

\bibitem[{{Bruzual} \& {Charlot}(2003)}]{Bruzual2003}
{Bruzual}, G. \& {Charlot}, S. 2003, \mnras, 344, 1000

\bibitem[{{Cai} {et~al.}(2015){Cai}, {Fan}, {Jiang}, {Dav{\'e}}, {Oh}, {Yang},
  \& {Zabludoff}}]{Cai2015}
{Cai}, Z., {Fan}, X., {Jiang}, L., {et~al.} 2015, \apjl, 799, L19

\bibitem[{{Calzetti} {et~al.}(2000){Calzetti}, {Armus}, {Bohlin}, {Kinney},
  {Koornneef}, \& {Storchi-Bergmann}}]{Calzetti2000}
{Calzetti}, D., {Armus}, L., {Bohlin}, R.~C., {et~al.} 2000, \apj, 533, 682

 \bibitem[{{Caruana} {et~al.}(2012){Caruana}, {Bunker}, {Wilkins}, {Stanway},
   {Lacy}, {Jarvis}, {Lorenzoni}, \& {Hickey}}]{Caruana2012}
 {Caruana}, J., {Bunker}, A.~J., {Wilkins}, S.~M., {et~al.} 2012, \mnras, 427,
   3055

\bibitem[{{Castellano} {et~al.}(2010{\natexlab{a}}){Castellano}, {Fontana},
  {Boutsia}, {Grazian}, {Pentericci}, {Bouwens}, {Dickinson}, {Giavalisco},
  {Santini}, {Cristiani}, {Fiore}, {Gallozzi}, {Giallongo}, {Maiolino},
  {Mannucci}, {Menci}, {Moorwood}, {Nonino}, {Paris}, {Renzini}, {Rosati},
  {Salimbeni}, {Testa}, \& {Vanzella}}]{Castellano2010}
{Castellano}, M., {Fontana}, A., {Boutsia}, K., {et~al.} 2010{\natexlab{a}},
  \aap, 511, A20+

\bibitem[{{Castellano} {et~al.}(2012){Castellano}, {Fontana}, {Grazian},
  {Pentericci}, {Santini}, {Koekemoer}, {Cristiani}, {Galametz}, {Gallerani},
  {Vanzella}, {Boutsia}, {Gallozzi}, {Giallongo}, {Maiolino}, {Menci}, \&
  {Paris}}]{Castellano2012}
{Castellano}, M., {Fontana}, A., {Grazian}, A., {et~al.} 2012, \aap, 540, A39

\bibitem[{{Castellano} {et~al.}(2010{\natexlab{b}}){Castellano}, {Fontana},
  {Paris}, {Grazian}, {Pentericci}, {Boutsia}, {Santini}, {Testa}, {Dickinson},
  {Giavalisco}, {Bouwens}, {Cuby}, {Mannucci}, {Cl{\'e}ment}, {Cristiani},
  {Fiore}, {Gallozzi}, {Giallongo}, {Maiolino}, {Menci}, {Moorwood}, {Nonino},
  {Renzini}, {Rosati}, {Salimbeni}, \& {Vanzella}}]{Castellano2010b}
{Castellano}, M., {Fontana}, A., {Paris}, D., {et~al.} 2010{\natexlab{b}},
  \aap, 524, A28

\bibitem[{{Castellano} {et~al.}(2014){Castellano}, {Sommariva}, {Fontana},
  {Pentericci}, {Santini}, {Grazian}, {Amorin}, {Donley}, {Dunlop}, {Ferguson},
  {Fiore}, {Galametz}, {Giallongo}, {Guo}, {Huang}, {Koekemoer}, {Maiolino},
  {McLure}, {Paris}, {Schaerer}, {Troncoso}, \& {Vanzella}}]{Castellano2014}
{Castellano}, M., {Sommariva}, V., {Fontana}, A., {et~al.} 2014, \aap, 566, A19

\bibitem[{{Dayal} {et~al.}(2009){Dayal}, {Ferrara}, {Saro}, {Salvaterra},
  {Borgani}, \& {Tornatore}}]{Dayal2009}
{Dayal}, P., {Ferrara}, A., {Saro}, A., {et~al.} 2009, \mnras, 400, 2000

%\bibitem[{{Dijkstra}(2015)}]{Dijkstra2015} 
 %{Dijkstra}, M. 2015, ArXiv e-prints [\eprint[arXiv]{1511.01218}]

\bibitem[{{Dijkstra}(2015)}]{Dijkstra2015} 
 {Dijkstra}, M. 2015, ArXiv:1511.01218

\bibitem[{{Finkelstein} {et~al.}(2015){Finkelstein}, {Ryan}, {Papovich},
  {Dickinson}, {Song}, {Somerville}, {Ferguson}, {Salmon}, {Giavalisco},
  {Koekemoer}, {Ashby}, {Behroozi}, {Castellano}, {Dunlop}, {Faber}, {Fazio},
  {Fontana}, {Grogin}, {Hathi}, {Jaacks}, {Kocevski}, {Livermore}, {McLure},
  {Merlin}, {Mobasher}, {Newman}, {Rafelski}, {Tilvi}, \&
  {Willner}}]{Finkelstein2015}
{Finkelstein}, S.~L., {Ryan}, Jr., R.~E., {Papovich}, C., {et~al.} 2015, \apj,
  810, 71

\bibitem[{{Fontana} {et~al.}(2000){Fontana}, {D'Odorico}, {Poli}, {Giallongo},
  {Arnouts}, {Cristiani}, {Moorwood}, \& {Saracco}}]{Fontana2000}
{Fontana}, A., {D'Odorico}, S., {Poli}, F., {et~al.} 2000, \aj, 120, 2206

\bibitem[{{Fontana} {et~al.}(2014){Fontana}, {Dunlop}, {Paris}, {Targett},
  {Boutsia}, {Castellano}, {Galametz}, {Grazian}, {McLure}, {Merlin},
  {Pentericci}, {Wuyts}, {Almaini}, {Caputi}, {Chary}, {Cirasuolo},
  {Conselice}, {Cooray}, {Daddi}, {Dickinson}, {Faber}, {Fazio}, {Ferguson},
  {Giallongo}, {Giavalisco}, {Grogin}, {Hathi}, {Koekemoer}, {Koo}, {Lucas},
  {Nonino}, {Rix}, {Renzini}, {Rosario}, {Santini}, {Scarlata}, {Sommariva},
  {Stark}, {van der Wel}, {Vanzella}, {Wild}, {Yan}, \&
  {Zibetti}}]{Fontana2014}
{Fontana}, A., {Dunlop}, J.~S., {Paris}, D., {et~al.} 2014, \aap, 570, A11

\bibitem[{{Fontana} {et~al.}(2010){Fontana}, {Vanzella}, {Pentericci},
  {Castellano}, {Giavalisco}, {Grazian}, {Boutsia}, {Cristiani}, {Dickinson},
  {Giallongo}, {Maiolino}, {Moorwood}, \& {Santini}}]{Fontana2010}
{Fontana}, A., {Vanzella}, E., {Pentericci}, L., {et~al.} 2010, \apjl, 725,
  L205

\bibitem[{{Gonzaga} \& {et al.}(2012)}]{Gonzaga2012}
{Gonzaga}, S. \& {et al.} 2012, {The DrizzlePac Handbook}

\bibitem[{{Grazian} {et~al.}(2012){Grazian}, {Castellano}, {Fontana},
  {Pentericci}, {Dunlop}, {McLure}, {Koekemoer}, {Dickinson}, {Faber},
  {Ferguson}, {Galametz}, {Giavalisco}, {Grogin}, {Hathi}, {Kocevski}, {Lai},
  {Newman}, \& {Vanzella}}]{Grazian2012}
{Grazian}, A., {Castellano}, M., {Fontana}, A., {et~al.} 2012, \aap, 547, A51

\bibitem[{{Guo} {et~al.}(2013){Guo}, {Ferguson}, {Giavalisco}, {Barro},
  {Willner}, {Ashby}, {Dahlen}, {Donley}, {Faber}, {Fontana}, {Galametz},
  {Grazian}, {Huang}, {Kocevski}, {Koekemoer}, {Koo}, {McGrath}, {Peth},
  {Salvato}, {Wuyts}, {Castellano}, {Cooray}, {Dickinson}, {Dunlop}, {Fazio},
  {Gardner}, {Gawiser}, {Grogin}, {Hathi}, {Hsu}, {Lee}, {Lucas}, {Mobasher},
  {Nandra}, {Newman}, \& {van der Wel}}]{Guo2013}
{Guo}, Y., {Ferguson}, H.~C., {Giavalisco}, M., {et~al.} 2013, \apjs, 207, 24

\bibitem[{{Hutter} {et~al.}(2015){Hutter}, {Dayal}, \&
  {M{\"u}ller}}]{hutter2015}
{Hutter}, A., {Dayal}, P., \& {M{\"u}ller}, V. 2015, \mnras, 450, 4025

\bibitem[{{Hutter} {et~al.}(2014){Hutter}, {Dayal}, {Partl}, \&
  {M{\"u}ller}}]{hutter2014}
{Hutter}, A., {Dayal}, P., {Partl}, A.~M., \& {M{\"u}ller}, V. 2014, \mnras,
  441, 2861

\bibitem[{{Iliev} {et~al.}(2014){Iliev}, {Mellema}, {Ahn}, {Shapiro}, {Mao}, \&
  {Pen}}]{Iliev2014}
{Iliev}, I.~T., {Mellema}, G., {Ahn}, K., {et~al.} 2014, \mnras, 439, 725

\bibitem[{{Loeb} {et~al.}(2005){Loeb}, {Barkana}, \& {Hernquist}}]{Loeb2005}
{Loeb}, A., {Barkana}, R., \& {Hernquist}, L. 2005, \apj, 620, 553

\bibitem[{{Koekemoer} {et~al.}(2013){Koekemoer}, {Ellis}, {McLure}, {Dunlop},
  {Robertson}, {Ono}, {Schenker}, {Ouchi}, {Bowler}, {Rogers}, {Curtis-Lake},
  {Schneider}, {Charlot}, {Stark}, {Furlanetto}, {Cirasuolo}, {Wild}, \&
  {Targett}}]{Koekemoer2013}
{Koekemoer}, A.~M., {Ellis}, R.~S., {McLure}, R.~J., {et~al.} 2013, \apjs, 209,
  3

\bibitem[{{McQuinn} {et~al.}(2007){McQuinn}, {Lidz}, {Zahn}, {Dutta},
  {Hernquist}, \& {Zaldarriaga}}]{McQuinn2007}
{McQuinn}, M., {Lidz}, A., {Zahn}, O., {et~al.} 2007, \mnras, 377, 1043

\bibitem[{{Merlin} {et~al.}(2015){Merlin}, {Fontana}, {Ferguson}, {Dunlop},
  {Elbaz}, {Bourne}, {Bruce}, {Buitrago}, {Castellano}, {Schreiber}, {Grazian},
  {McLure}, {Okumura}, {Shu}, {Wang}, {Amor{\'{\i}}n}, {Boutsia}, {Cappelluti},
  {Comastri}, {Derriere}, {Faber}, \& {Santini}}]{Merlin2015}
{Merlin}, E., {Fontana}, A., {Ferguson}, H.~C., {et~al.} 2015, \aap, 582, A15

\bibitem[{{Mitra} {et~al.}(2015){Mitra}, {Choudhury}, \& {Ferrara}}]{Mitra2015}
{Mitra}, S., {Choudhury}, T.~R., \& {Ferrara}, A. 2015, \mnras, 454, L76

% \bibitem[{{Momcheva} {et~al.}(2015){Momcheva}, {Brammer}, {van Dokkum},
%   {Skelton}, {Whitaker}, {Nelson}, {Fumagalli}, {Maseda}, {Leja}, {Franx},
%   {Rix}, {Bezanson}, {Da Cunha}, {Dickey}, {F{\"o}rster Schreiber},
%   {Illingworth}, {Kriek}, {Labb{\'e}}, {Ulf Lange}, {Lundgren}, {Magee},
%   {Marchesini}, {Oesch}, {Pacifici}, {Patel}, {Price}, {Tal}, {Wake}, {van der
%   Wel}, \& {Wuyts}}]{Momcheva2015}
% {Momcheva}, I.~G., {Brammer}, G.~B., {van Dokkum}, P.~G., {et~al.} 2015, ArXiv
%   e-prints [\eprint[arXiv]{1510.02106}]

\bibitem[{{Momcheva} {et~al.}(2015){Momcheva}, {Brammer}, {van Dokkum},
  {Skelton}, {Whitaker}, {Nelson}, {Fumagalli}, {Maseda}, {Leja}, {Franx},
  {Rix}, {Bezanson}, {Da Cunha}, {Dickey}, {F{\"o}rster Schreiber},
  {Illingworth}, {Kriek}, {Labb{\'e}}, {Ulf Lange}, {Lundgren}, {Magee},
  {Marchesini}, {Oesch}, {Pacifici}, {Patel}, {Price}, {Tal}, {Wake}, {van der
  Wel}, \& {Wuyts}}]{Momcheva2015}
{Momcheva}, I.~G., {Brammer}, G.~B., {van Dokkum}, P.~G., {et~al.} 2015, arXiv:1510.02106

\bibitem[{{Ono} {et~al.}(2012){Ono}, {Ouchi}, {Mobasher}, {Dickinson},
  {Penner}, {Shimasaku}, {Weiner}, {Kartaltepe}, {Nakajima}, {Nayyeri},
  {Stern}, {Kashikawa}, \& {Spinrad}}]{Ono2012}
{Ono}, Y., {Ouchi}, M., {Mobasher}, B., {et~al.} 2012, \apj, 744, 83

\bibitem[{{Partl} {et~al.}(2011){Partl}, {Maselli}, {Ciardi}, {Ferrara}, \&
  {M{\"u}ller}}]{Partl2011}
{Partl}, A.~M., {Maselli}, A., {Ciardi}, B., {Ferrara}, A., \& {M{\"u}ller}, V.
  2011, \mnras, 414, 428

\bibitem[{{Pentericci} {et~al.}(2011){Pentericci}, {Fontana}, {Vanzella},
  {Castellano}, {Grazian}, {Dijkstra}, {Boutsia}, {Cristiani}, {Dickinson},
  {Giallongo}, {Giavalisco}, {Maiolino}, {Moorwood}, {Paris}, \&
  {Santini}}]{Pentericci2011}
{Pentericci}, L., {Fontana}, A., {Vanzella}, E., {et~al.} 2011, \apj, 743, 132

\bibitem[{{Pentericci} {et~al.}(2014){Pentericci}, {Vanzella}, {Fontana},
  {Castellano}, {Treu}, {Mesinger}, {Dijkstra}, {Grazian}, {Brada{\v c}},
  {Conselice}, {Cristiani}, {Dunlop}, {Galametz}, {Giavalisco}, {Giallongo},
  {Koekemoer}, {McLure}, {Maiolino}, {Paris}, \& {Santini}}]{Pentericci2014}
{Pentericci}, L., {Vanzella}, E., {Fontana}, A., {et~al.} 2014, \apj, 793, 113

 \bibitem[{{Schaerer} \& {de Barros}(2009)}]{Schaerer2009}
 {Schaerer}, D. \& {de Barros}, S. 2009, \aap, 502, 423


\bibitem[{{Schenker} {et~al.}(2012){Schenker}, {Stark}, {Ellis}, {Robertson},
  {Dunlop}, {McLure}, {Kneib}, \& {Richard}}]{Schenker2012}
{Schenker}, M.~A., {Stark}, D.~P., {Ellis}, R.~S., {et~al.} 2012, \apj, 744,
  179

\bibitem[{{Springel}(2005)}]{Springel2005}
{Springel}, V. 2005, \mnras, 364, 1105

\bibitem[{{Stark} {et~al.}(2010){Stark}, {Ellis}, {Chiu}, {Ouchi}, \&
  {Bunker}}]{Stark2010}
{Stark}, D.~P., {Ellis}, R.~S., {Chiu}, K., {Ouchi}, M., \& {Bunker}, A. 2010,
  \mnras, 408, 1628

\bibitem[{{Trenti} \& {Stiavelli}(2008)}]{Trenti2008}
{Trenti}, M. \& {Stiavelli}, M. 2008, \apj, 676, 767

\bibitem[{{Treu} {et~al.}(2012){Treu}, {Trenti}, {Stiavelli}, {Auger}, \&
  {Bradley}}]{Treu2012}
{Treu}, T., {Trenti}, M., {Stiavelli}, M., {Auger}, M.~W., \& {Bradley}, L.~D.
  2012, \apj, 747, 27


\bibitem[{{Vanzella} {et~al.}(2014){Vanzella}, {Fontana}, {Zitrin}, {Coe},
  {Bradley}, {Postman}, {Grazian}, {Castellano}, {Pentericci}, {Giavalisco},
  {Rosati}, {Nonino}, {Smit}, {Balestra}, {Bouwens}, {Cristiani}, {Giallongo},
  {Zheng}, {Infante}, {Cusano}, \& {Speziali}}]{Vanzella2014}
{Vanzella}, E., {Fontana}, A., {Zitrin}, A., {et~al.} 2014, \apjl, 783, L12

\bibitem[{{Vanzella} {et~al.}(2011){Vanzella}, {Pentericci}, {Fontana},
  {Grazian}, {Castellano}, {Boutsia}, {Cristiani}, {Dickinson}, {Gallozzi},
  {Giallongo}, {Giavalisco}, {Maiolino}, {Moorwood}, {Paris}, \&
  {Santini}}]{Vanzella2011}
{Vanzella}, E., {Pentericci}, L., {Fontana}, A., {et~al.} 2011, \apjl, 730,
  L35+

\bibitem[{{Wyithe} \& {Loeb}(2005)}]{Wyithe2005}
{Wyithe}, J.~S.~B. \& {Loeb}, A. 2005, \apj, 625, 1

\bibitem[{{Wyithe} \& {Loeb}(2007)}]{Wyithe2007}
{Wyithe}, J.~S.~B. \& {Loeb}, A. 2007, \mnras, 382, 921

\end{thebibliography}
\end{document}